\documentclass[aps,prl,superscriptaddress,twocolumn,10pt]{revtex4-1}
\usepackage{graphicx}
\usepackage{physics}
\usepackage{algpseudocode}
\usepackage{pythonhighlight}
\usepackage{amsmath,amssymb,amsfonts}
\usepackage{hyperref}
\usepackage{color}
\usepackage{natbib}
\usepackage{changes}
\usepackage[english]{babel}
\usepackage{subcaption}
\usepackage{subfiles}
\usepackage{float}
\usepackage{siunitx}
\usepackage{mhchem}
\AtBeginDocument{\RenewCommandCopy\qty\SI}

\begin{document}

\title{Towards quantum utility for NMR quantum simulation on a NISQ computer}
\author{Artemiy Burov}
\affiliation{School of Life Sciences, University of Applied Sciences Northwestern Switzerland (FHNW), Hofackerstrasse 30, CH-4132 Muttenz, Switzerland}
\affiliation{\'{E}cole Polytechnique F\'{e}d\'{e}rale de Lausanne (EPFL), CH-1015 Lausanne, Switzerland}
\affiliation{MatterDecoder, Lausanne, CH-1015, Switzerland}
\author{Oliver Nagl}
\affiliation{School of Life Sciences, University of Applied Sciences Northwestern Switzerland (FHNW), Hofackerstrasse 30, CH-4132 Muttenz, Switzerland}
\affiliation{ETH Zurich, CH-8093 Zurich, Switzerland}
\author{Cl\'{e}ment Javerzac-Galy}
\email{clement.javerzac@fhnw.ch}
\affiliation{School of Life Sciences, University of Applied Sciences Northwestern Switzerland (FHNW), Hofackerstrasse 30, CH-4132 Muttenz, Switzerland}
\affiliation{MatterDecoder, Lausanne, CH-1015, Switzerland}

\begin{abstract}
While the recent demonstration of accurate computations of classically intractable simulations on noisy quantum processors brings quantum advantage closer, there is still the challenge of demonstrating it for practical problems. Here we investigate the application of noisy intermediate-scale quantum devices for simulating nuclear magnetic resonance (NMR) experiments in the high-field regime. In this work, the NMR interactions are mapped to a quantum device via a product formula with minimal resource overhead, an approach that we discuss in detail. Using this approach, we show the results of simulations of liquid-state proton NMR spectra on relevant molecules with up to $11$ spins, and up to a total of $47$ atoms, and compare them with real NMR experiments. Despite current limitations, we show that a similar approach will eventually lead to a case of \emph{quantum utility}, a scenario where a practically relevant computational problem can be solved by a quantum computer but not by conventional means. We provide an experimental estimation of the amount of quantum resources needed for solving larger instances of the problem with the presented approach. The polynomial scaling we demonstrate on real processors is a foundational step in bringing practical quantum computation closer to reality.
\end{abstract}

\maketitle

\section{Introduction}

The demonstration of a useful application of the noisy intermediate-scale quantum (NISQ) technology is of substantial interest to the quantum technology industry. It is still an open question whether the utility of quantum computing can be achieved in the pre-fault-tolerant era. In this work, we show a path towards this possibility. As recently outlined by Kim, Eddins \textit{et al.} \cite{Kim2023}, quantum advantage can be approached in two ways. Firstly, by engineering artificial classically hard problems that can be efficiently solved on a quantum device, as was demonstrated in \cite{Kim2023}. Secondly, by finding practical problems with the same requirement: classically hard problems that can be efficiently solved on a quantum device. While the first way has been demonstrated, albeit it was recently shown that the exact problem statement considered in \cite{Kim2023} is in fact not classically sufficiently hard to construct a bulletproof argument \cite{Begusic2024-ny}, we focus on the second one.

\begin{figure*}[hbtp!]
    \centering
    \includegraphics[width=0.8\textwidth]{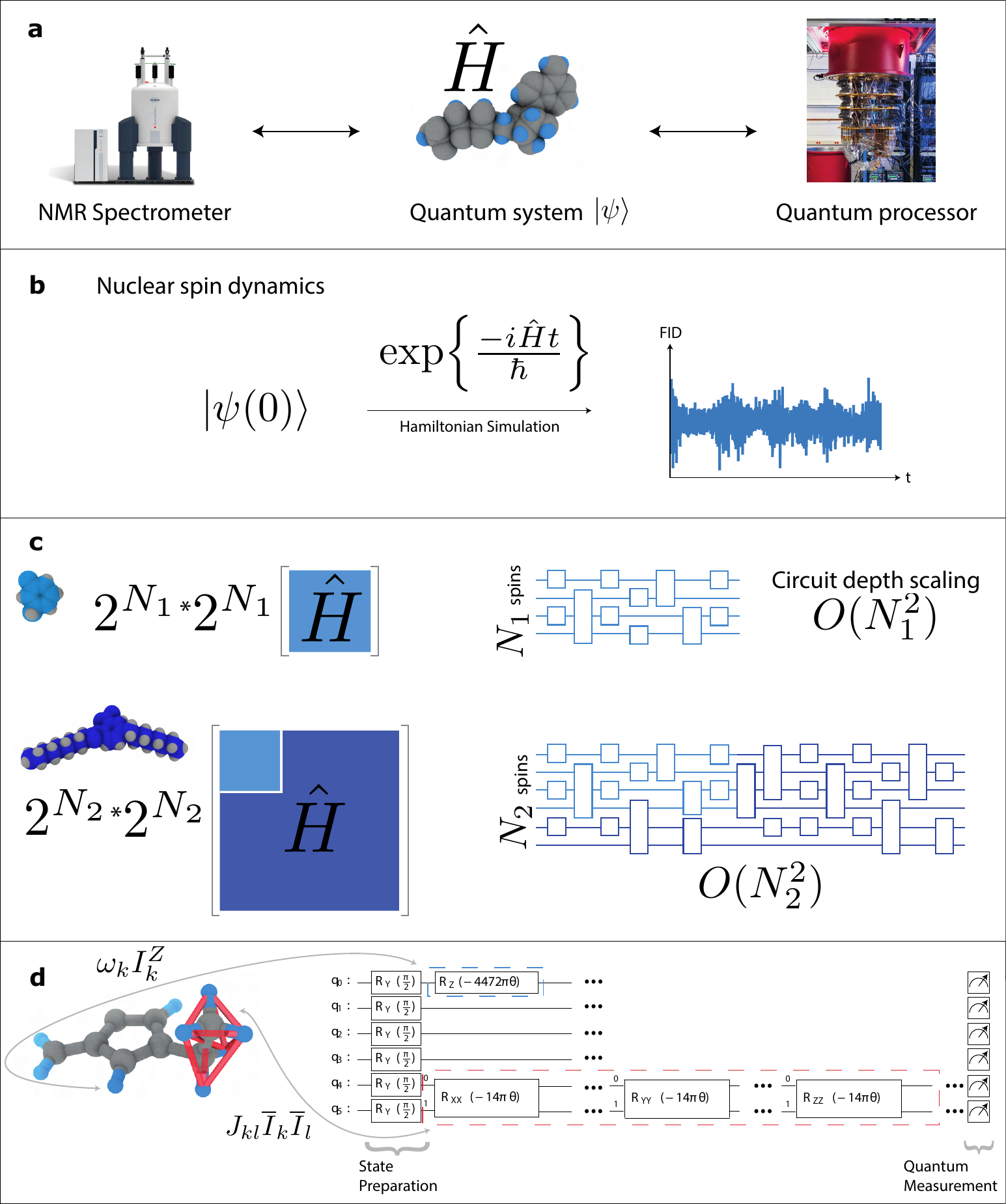}
    \caption{(a) shows the main characters of this work: either the NMR spectrometer or the quantum processor generates spectra of a given quantum spin system. (b) shows the pathway of NMR Hamiltonian simulation (quantum or classical). The signal is calculated for a set of time points by applying the time propagator $\exp{\frac{-i\hat{H}t}{\hbar}}$ to the initial state $\ket{\psi(0)}$. The propagator can be applied either analytically or by a quantum calculation. (c) shows the scaling of various mathematical objects involved in the discussion. The Hamiltonian scales exponentially - indicating the need for an exponential amount of classical resources required for the simulation. Required quantum resources scale only polynomially since the product formula quantum circuit scales polynomially. (d) shows a part of the mapping between a molecule (3-Amino-5-ethyl-1H-pyrazole) and its quantum circuit for the FID signal calculation. The spin-field couplings are represented by single qubit rotations around the $Z$ axis, and the spin-spin couplings are represented by rotations around $XX$, $YY$, and $ZZ$ axes of the corresponding two-qubit Hilbert space.}
    \label{circuit}
\end{figure*}

Here we discuss the problem of Heisenberg Hamiltonian simulation, an important building block for a set of problems revolving around spin dynamics of \textit{nuclear} magnetic resonance (NMR) (figure \ref{circuit} \textbf{a} and \textbf{b}) previously discussed in the context of zero-field NMR \cite{demler1, demler2, google_obrian}. Compared to some other quantum chemistry tasks this application offers a particularly favorable polynomial scaling, as illustrated in figure \ref{circuit} \textbf{c}. Simulation of \textit{fermionic} systems described by second quantization Hamiltonians is interesting in particular for understanding the mechanisms behind high-energy superconductivity or in many \textit{electronic} quantum chemistry problems but requires a costly encoding (for example the Bravyi-Kitaev transformation) to be mapped to a quantum circuit. The task of finding low energy eigenvalues of molecules is another practically interesting problem statement, but this is a \emph{BQNP-hard} problem. BQNP (Bounded-error Quantum Non-deterministic Polynomial-time, closely related to well-known QMA) is the quantum counterpart of the classical NP complexity class as described in \cite{kitaev2002classical}, and resources required for finding solutions to such BQNP-hard problems scale superpolynomially with the size of the instance. This type of quantum computation is to our knowledge beyond the capabilities of quantum computers available today and in the next few years for practically interesting problem instances.

In the proposed \textit{nuclear} magnetic spin dynamics simulations, the objects of interest are spatially localized fermions, which can be described with similar mathematics as qubits. This removes any encoding overhead.

Furthermore, NMR is in fact the grandfather of quantum computers, ultimately going into early retirement as a quantum computing architecture facing scalability and noise issues, but leaving us with a whole array of spin manipulation techniques that are used in quantum computing today \cite{RevModPhys.76.1037}. It is no surprise that a system that can be used as a quantum computer is itself well suited to be simulated by a quantum computer.

Apart from this conceited interest from the quantum computing community, we believe that the present set of problems is also of profound interest to the NMR community, in particular for applications such as chemical structure verification and elucidation via NMR spectroscopy. NMR spectroscopy is widely used in many fields of science and industry \cite{japan2017experimental}, such as organic chemistry, medicine, agriculture, food chemistry, and environmental sciences. This work could also ultimately impact computationally hard NMR problems in catalyst and battery development \cite{Reif2021}.

As experience and previous work \cite{demler1, demler2, google_obrian} show, the simulation of NMR spectra is a crucial building block for these practical applications. It can benefit from quantum information processing, since this problem is equivalent to simulating the action of a Hamiltonian on a quantum system, which is believed to be a non-trivial problem, though among the easiest to address with early quantum devices \cite{childs2018}. The idea of simulating microscopic nature, a natively quantum problem, on a quantum machine, was one of the initial motivations behind the development of quantum computing \cite{Manin1980, Feynman1982}.

\begin{figure*}[ht!]
    \centering
    \includegraphics[width=\textwidth]{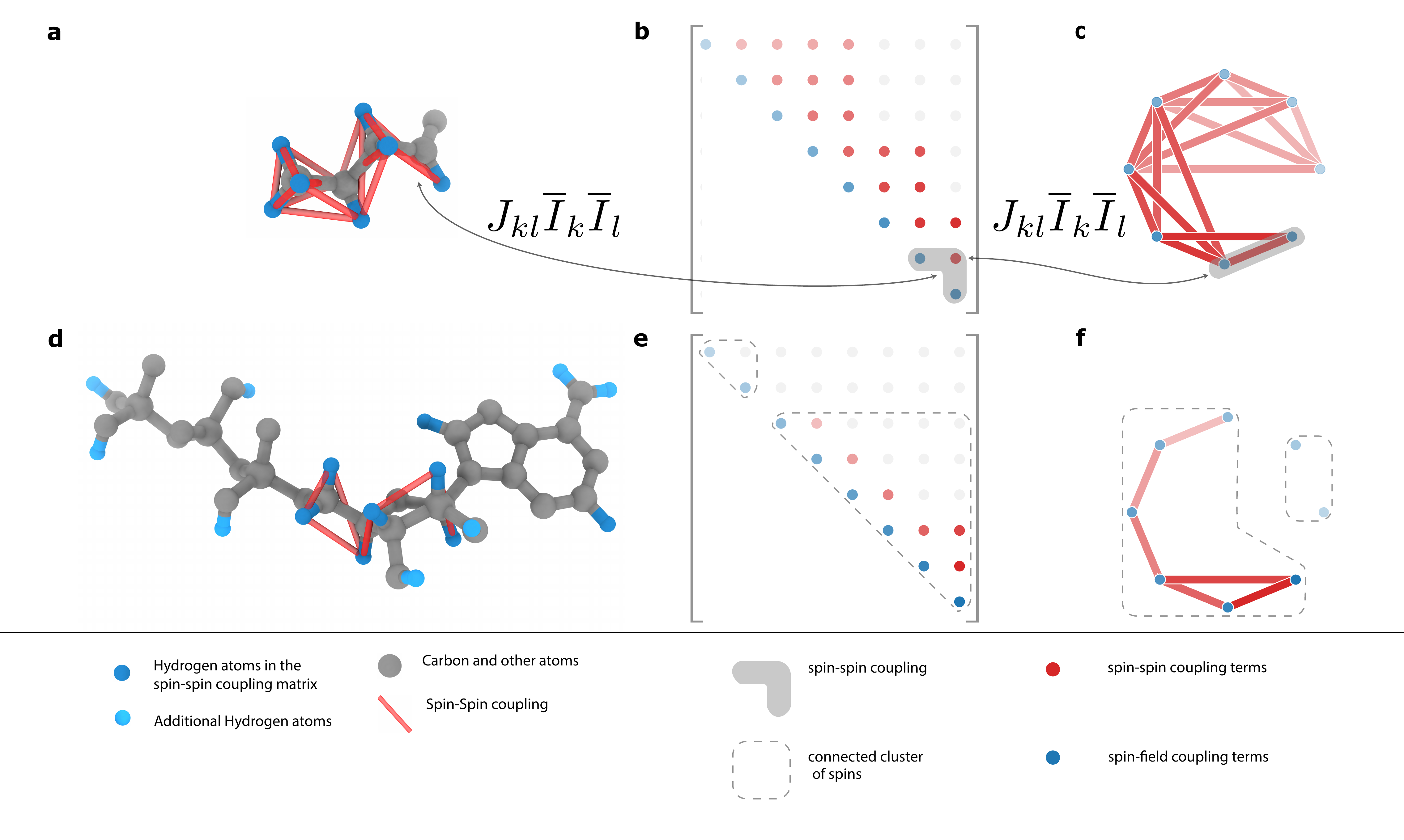}
    \caption{Illustrates the concepts of the spin-spin coupling matrix and the spin-spin coupling graph. The presented molecules are Butyraldehyde (a, b, and c) and Adenosine triphosphate (d, e, and f). The high connectivity of spins in Butyraldehyde can be seen by the number of connections in the corresponding graph c) and the number of off-diagonal elements in the matrix b). The number of off-diagonal elements is proportional to the number of \emph{two-qubit gates} needed for the simulation of the spin cluster, an essential resource in quantum computing. Block-diagonal matrices such as seen in e) indicate a spin system that is separable into independent subsystems, each with independent unitary dynamics.}
    \label{graph}
\end{figure*}

In this work, we show the simulation of high-field NMR free induction decay (FID) experiments on a quantum processor (QPU). To our knowledge, this is the first such simulation. We compare the results with the output of NMR experiments performed on a $\qty{400}{\MHz}$ instrument.

We first explain the principles of the NMR FID experiment and discuss cases when this problem becomes computationally complex. Without loss of generality, we only consider $^1$H hydrogen spins in this work, in which case one speaks of \emph{proton} NMR. The way to map this experiment on a QPU is then discussed. This includes the discussion of the quantum algorithm and the technicalities of implementing this algorithm on quantum hardware. We then show the results of the quantum simulation for some of the hardest cases that we can currently simulate with the proposed approach, around $10$ spins, depending on the structure of the molecule. The last results in \cite{demler2} showed simulations for a molecule containing a methyl group with $3$ hydrogen atoms that were probed in a zero-field experiment. We also cover some topics that may be discussed in more detail in future work. In particular, the resources needed for harder cases of the NMR FID simulation problem are covered.

\section{NMR Theoretical Background}

The NMR Hamiltonian describes the dynamics of the magnetic system of nuclear spins. Two types of interactions define these dynamics: the \emph{spin-field} interaction that gives rise to \emph{chemical shifts} in the NMR spectrum and the \emph{spin-spin} interaction that causes splittings in the chemical shifts - or at least such is the heuristic point of view that is commonly assumed in interpreting spectra obtained in \emph{high field} \emph{liquid-state} NMR experiments. In more general cases, such as \emph{near-zero field} NMR \cite{PhysRevLett.107.107601}, \emph{solid-state} NMR \cite{Reif2021}, and also in the case of \emph{strongly correlated} high-field liquid-state NMR \cite{nmr_coupling}, this heuristic point of view and the corresponding heuristic methods of NMR spectra analysis break down.

It is in general necessary to calculate the action of the Hamiltonian on the Hilbert space of the spin system to get information about its dynamics. This calculation is computationally hard - in general, it scales exponentially with the size of the spin system, as does the Hamiltonian matrix, see figure \ref{circuit} \textbf{c}. It is mathematically equivalent to applying an exponential of a $2^N$ by $2^N$ sparse matrix to a vector, where $N$ is the size of the spin system. In many cases, numerical methods can be employed to perform this computation efficiently \cite{kuprovspin}. This is especially true for the cases where $N$ is small or the spins in the molecule are weakly connected.

In the most difficult cases, such as molecules with highly interconnected spins, the numerical methods fail \cite{kuprov_karabanov}. A good example where exponential complexity is unavoidable is the NMR of single crystals, famously calcium fluoride \cite{cf2}. For this reason, the Hamiltonian simulation was proposed as a solution \cite{demler1, demler2, google_obrian}.

\paragraph{\textbf{Spin connectivity.}}

To visualize the connectivity of the molecule, it helps to introduce the concepts of the spin-spin coupling matrix and the spin-spin coupling graph. In figure \ref{graph}, two molecules are presented as examples, both with $8$ $^1$H atoms in the considered spin systems.

The first example, Butyraldehyde, is a relatively small molecule with a highly connected spin system. This can be seen in the number of off-diagonal elements in the spin-spin coupling matrix, each of them corresponds to a spin-spin coupling. It also can be seen in the highly connected spin-spin coupling graph.

The second molecule, adenosine triphosphate (ATP), has more $^1$H atoms than the $8$ in the shown spin system. Often, a spin will be decoupled from other spins, for example, due to being in a bond with an oxygen atom. This results in a $^1$H atom heavily shielded from its environment, which frequently exchanges with solvent $^1$H atoms at higher temperatures \cite{exarchou2002strong}. The dynamics of the whole system can be considered as the set of separate unitary evolutions of non-interacting subsystems of the whole system. For the NMR spectra, it effectively means that they can be produced as a sum of the spectra of all the non-interacting (or weakly interacting) subsystems of the whole spin system - the spin clusters.

Two of the non-interacting atoms are included in the considered spin system. This can be seen in the spin-spin coupling matrix, as it is block-diagonalizable. The blocks in this representation correspond to the interacting subsystems. It also can be seen in the disconnected spin-spin coupling graph.

When the spin system together with its interactions is mapped to a quantum circuit, the amount of the most expensive computational resource needed (in terms of noise) - the two-qubit quantum gates - is directly proportional to the number of the non-zero off-diagonal elements. In the worst-case scenario of the fully connected spin system, the number of the two-qubit quantum gates scales as the number of elements in the upper half of the spin-spin coupling matrix, so it is quadratic in the size $N$ of the spin system. This is rare in liquid-state NMR.

\paragraph{\textbf{NMR FID experiment.}}

In the NMR FID experiment, the signal is produced by the magnetization of nuclear spins of a molecular system of interest precessing around a magnetic field \cite{levitt}. Usually, only one species of nuclei is excited, for example, the $^1$H of the molecule in the case of proton NMR, as considered in this work without loss of generality.

This FID signal allows us to gain information about the structure of the studied sample. Mathematically the structure of the sample is contained in the Hamiltonian $\hat{H}$ of the studied system. The Hamiltonian also defines the dynamics of the studied system via the Schr\"odinger's equation, provided for the case of a pure state $ \ket{\psi} $ by 
\begin{equation}\label{schroe}
    i\hbar\dv{}{t}\ket{\psi(t)}=\hat{H}\ket{\psi(t)}.
\end{equation} 
For the case of time-independent Hamiltonian, the formal solution for the time evolution can be written simply as 
\begin{equation}\label{timeevo}
    \ket{\psi(t)}=\exp{\frac{-i\hat{H} t}{\hbar}} \ket{\psi(0)}.
\end{equation}
The exponential of the Hamiltonian matrix in eq. (\ref{timeevo}) is hard to compute. It is this computational barrier that hopefully can be traversed using quantum computing. 

Once the state $\ket{\psi(t)}$ is stored in a register of a quantum computer, a projective quantum measurement of the total magnetization observable $\hat{M}_X = \sum^N_{k=1} I^X_k$ can be performed to obtain the value of magnetization at time $t$, where $I^X_k$ is the $X$ component of the spin operator for the $k$-th spin. The measurement collapses the quantum state into one of the eigenspaces of the observable, and the information about its previous evolution is lost. For this reason, the evolution of the state has to be performed for each recorded time point of the simulated FID. This seemingly large overhead is however only constant, that is, it does not depend on the size of the simulated spin system.

The NMR Hamiltonian is often described as a Heisenberg Hamiltonian \cite{levitt}. It is commonly written as
\begin{equation}\label{NMRH1}
     \hat{H}=-\sum^N_{k=1} \gamma \overline{I}_k(1-\sigma_k)\overline{B}_0 + 2\pi\sum_{k<l}\overline{I}_kJ_{kl}\overline{I}_l
\end{equation}
for a liquid-state NMR experiment with spin $\frac{1}{2}$ nuclei, which is the case considered in this work. Term-by term it can be understood as $\gamma\overline{I}_k\overline{B}_0=\omega_0\overline{I}_k$ describes the interaction of unshielded nucleus with the magnetic field, where $\gamma$ is the gyromagnetic ratio of the studied nuclei, hydrogens in our case. $\overline{I}_k$ is the vector of spin operators. $\overline{B}_0$ is the magnetic field vector. $-\gamma\overline{I}_k\sigma_k\overline{B}_0$ describes the modification to this interaction due to the shielding by the electrons of the molecule, the \emph{chemical shielding}, where $\sigma_k$ is the chemical shift. $2\pi\sum_{k<l}\overline{I}_kJ_{kl}\overline{I}_l$ describes the spin-spin interaction, where $J_{kl}$ is the J-coupling. We write $I_k$ here and in the following without the characteristic operator hat to avoid unnecessary cluttering of text. Nevertheless, it is to be understood as an operator, and $\overline{I}_k$ as a vector of operators. In general, both $\sigma_k$ and $J$ can be a tensor, in which case the interaction is called \emph{anisotropic}. In this work, we consider only isotropic interactions commonly encountered in liquid-state NMR. $\overline{I}_kJ_{kl}\overline{I}_l$ is a bilinear form, and in the isotropic case it can be rewritten simply as $J_{kl}\overline{I}_k\overline{I}_l$ since $J_{kl}$ is a scalar.

We will now rewrite this Hamiltonian in a format more suitable for the desired simulations. Without loss of generality, the field $\overline{B}_0$ can be considered to point along the $Z$ direction. Also, as in this work, we assume interactions to be isotropic, both $\sigma_k$ and $J$ are scalars. With these considerations, we can rewrite the Hamiltonian as
\begin{equation}\label{NMRH2}
\hat{H}=\sum^N_{k=1} \omega_{0k}I^Z_k + 2\pi\sum_{k<l}J_{kl}\overline{I}_k\overline{I}_l
\end{equation}
with \emph{shifted} Larmor frequency of $k$-th nucleus given by $\omega_{0k}$.

Finally, for many applications, including the simulation presented in this work, it helps to consider the Hamiltonian in the rotating frame, where $\omega_{k}$ now is the shifted Larmor frequency in the rotating frame:

\begin{equation}\label{NMRH3}
\hat{H}^r=\sum^N_{k=1} \omega_{k}I^Z_k + 2\pi\sum_{k<l}J_{kl}\overline{I}_k\overline{I}_l
\end{equation}
Indeed, since the generator of rotations around the $Z$ axis given by $\sum_i I^Z_i$ commutes with all terms containing the $Z$ angular momentum operators $I^Z_k$ of the Hamiltonian $\hat{H}$, and additionally that it also commutes with all other terms that come up in this Hamiltonian, 
\begin{equation}\label{commutation}
\begin{split}
[\sum_i I^Z_i, I^X_kI^X_l + I^Y_kI^Y_l] =
[I^Z_k + I^Z_l, I^X_kI^X_l + I^Y_kI^Y_l] = \\
[I^Z_k, I^X_kI^X_l] + [I^Z_l, I^X_kI^X_l] + [I^Z_k, I^Y_kI^Y_l] + [I^Z_l, \hat{\sigma}^Y_kI^Y_l] = \\
iI^Y_kI^X_l + iI^X_kI^Y_l - iI^X_kI^Y_l - iI^Y_kI^X_l = 0,
\end{split}
\end{equation}
\begin{equation}\label{commutation2}
\begin{split}
[\sum_i I^Z_i, I^Z_kI^Z_l] = 0
\end{split}   
\end{equation}
we conclude that the Hamiltonian is invariant under rotations around the $Z$ axis. Unsurprisingly, since the unitary evolutions of all connected subsystems of the spin system can be considered separately, the Hamiltonian of the full spin systems also commutes with a transformation that involves different rotations for each subsystem. This can also be seen from eq. (\ref{commutation}) by grouping on the right side of the commutator $[\sum_i I^Z_i, \hat{H}^r]$ all $2\pi\sum_{k<l}J_{kl}\overline{I}_k\overline{I}_l$ terms that belong to the same connected subsystem. In that case on the left side of the commutator, the whole sum over all spins is not necessary - it is enough to sum over the spins of that subcluster to get the commutator to be equal to $0$. This sum gives a generator of rotations of that subcluster, and following this procedure for other subclusters we receive a generator of rotations that commutes with the Hamiltonian of the full system for each subcluster.

In this work we consider the initial state of the spin system to be a pure state. The spins are initially aligned with the $Z$ axis and then turned $90$ degrees around the $Y$ axis to point along the $X$ axis with a $\frac{\pi}{2}$ pulse.

The FID signal is obtained as an ensemble measurement of the total magnetization operators along $X$ and $Y$ axes at different points in time. These operators can be written as $\hat{M}_X = \sum^N_{k=1} I^X_k$ and $\hat{M}_Y = \sum^N_{k=1} I^Y_k$, respectively.

\begin{figure*}[ht!]
    \centering
    \includegraphics[width=\textwidth]{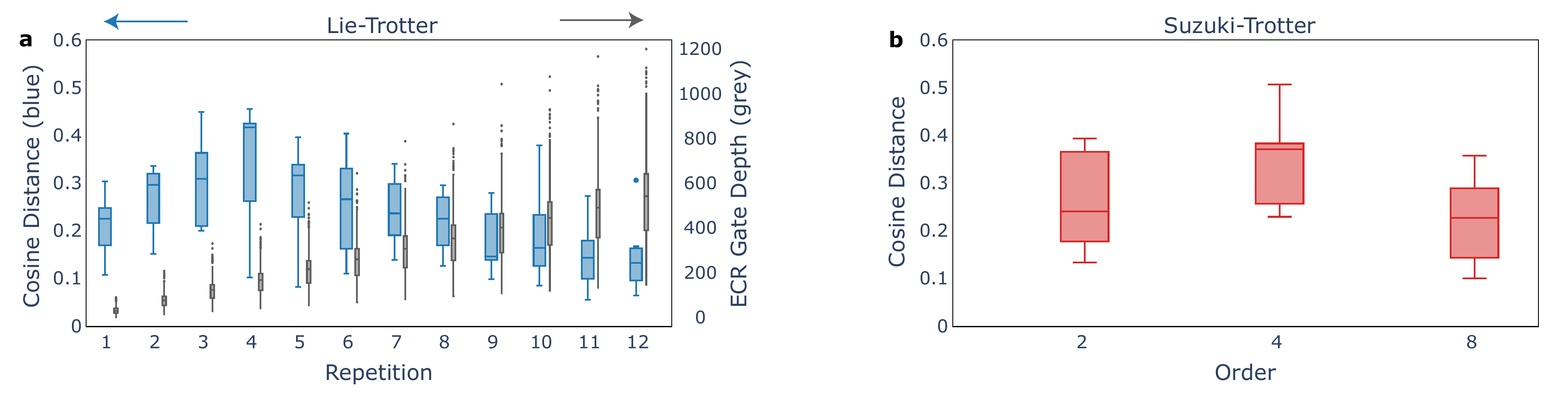}
    \caption{Shows the cosine distance between the spectra obtained by (a) Lie-Trotter and (b) Suzuki-Trotter product formulas and by exact Hamiltonian exponentiation methods. The calculation is done on $4$-spin molecules from the GISSMO \cite{gissmo} dataset. The grey bar plots in (a) show the scaling of the depth of two-qubit gates in the quantum circuit needed for the FID simulation, this can be thought of as the computational cost of the simulation.}
    \label{trotter}
\end{figure*}

\begin{figure*}[hbtp!]
    \centering
    \includegraphics[width=\textwidth]{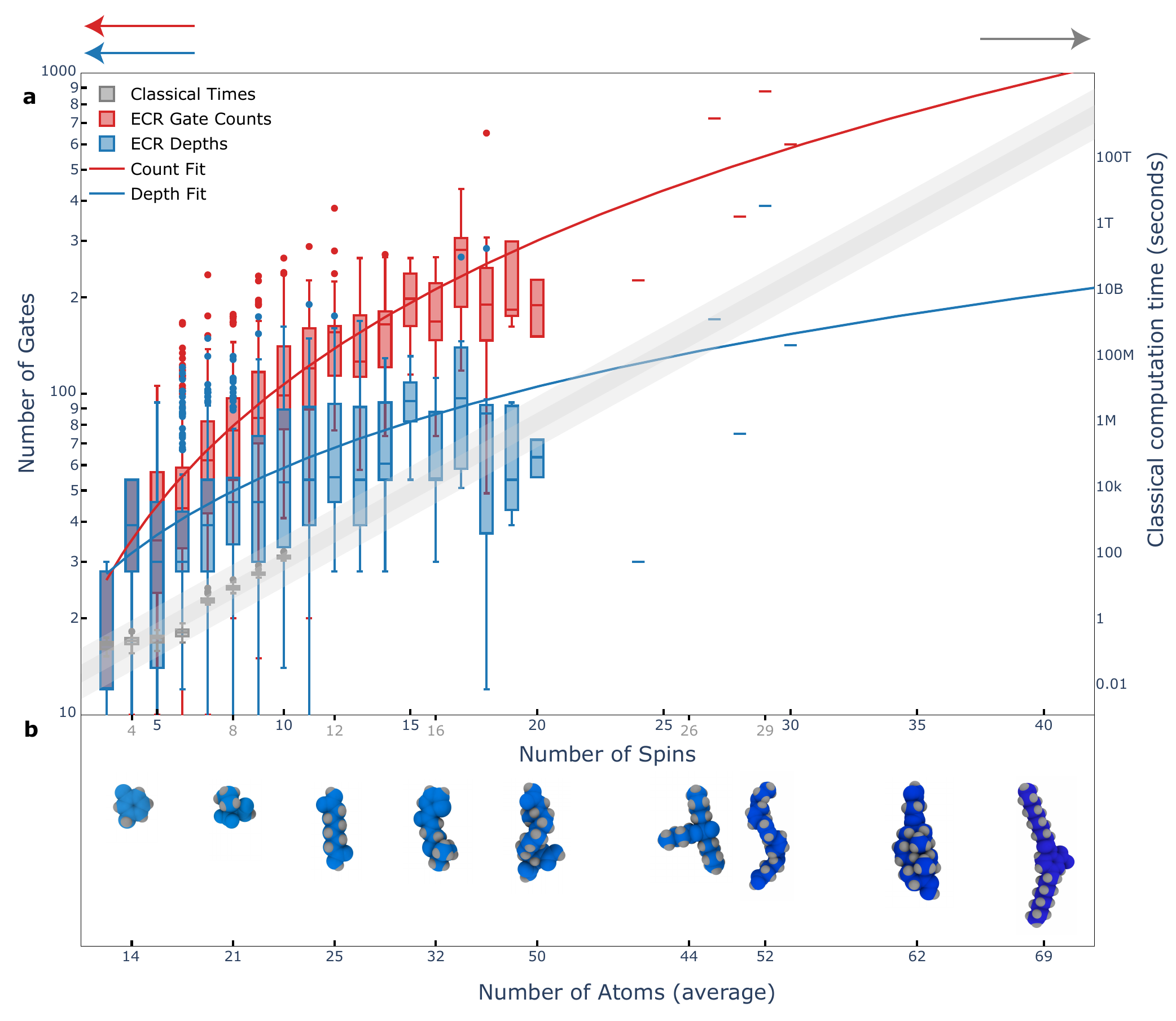}
    \caption{a) The left y-axis (red and blue data points, arrows pointing to the corresponding axis for visibility) shows the total number and depth of two-qubit echoed cross-resonance (ECR) gates needed to simulate a single repetition of a product formula for a given number of spins. The molecules are taken from the GISSMO database \cite{gissmo}. The quantum circuits are transpired to the IBM "brisbane" quantum device (Eagle r3). The transpilation algorithm is probabilistic, but in the considered cases the deviations in the number of gates for single molecules are low, in the order of $10$. The deviations in the box plots come mostly from the fact that the number of required gates depends on the connectivity of the molecule, and there are multiple molecules with the same number of hydrogen spins. In the extreme case where the molecule is not connected at all, the gate depth can be as low as $2$.
    The right y-axis (grey data points, arrow pointing to the corresponding axis for visibility) shows an estimation of the required runtime on a classical computer, calculated by taking the matrix exponential of the Hamiltonian, as a visual guide. We want to stress that for weakly correlated spin clusters these computations can be sped up by employing numerical methods to take the exponential of the Hamiltonian matrix, but in general cases, the computational complexity of this task scales up exponentially (see figure \ref{circuit}).
    b) Examples of molecules with $^1$H atoms highlighted in blue. These molecules are chosen as representative examples with the average number of atoms for a given number of spins, taken from the GISSMO database \cite{gissmo}. Molecules considered even in this regime with low numbers of hydrogen atoms can get quite large. Data for larger molecules is, at least in the GISSMO database, scarce.}
    \label{gates}
\end{figure*}

\begin{figure*}[ht!]
    \centering
    \includegraphics[width=\textwidth]{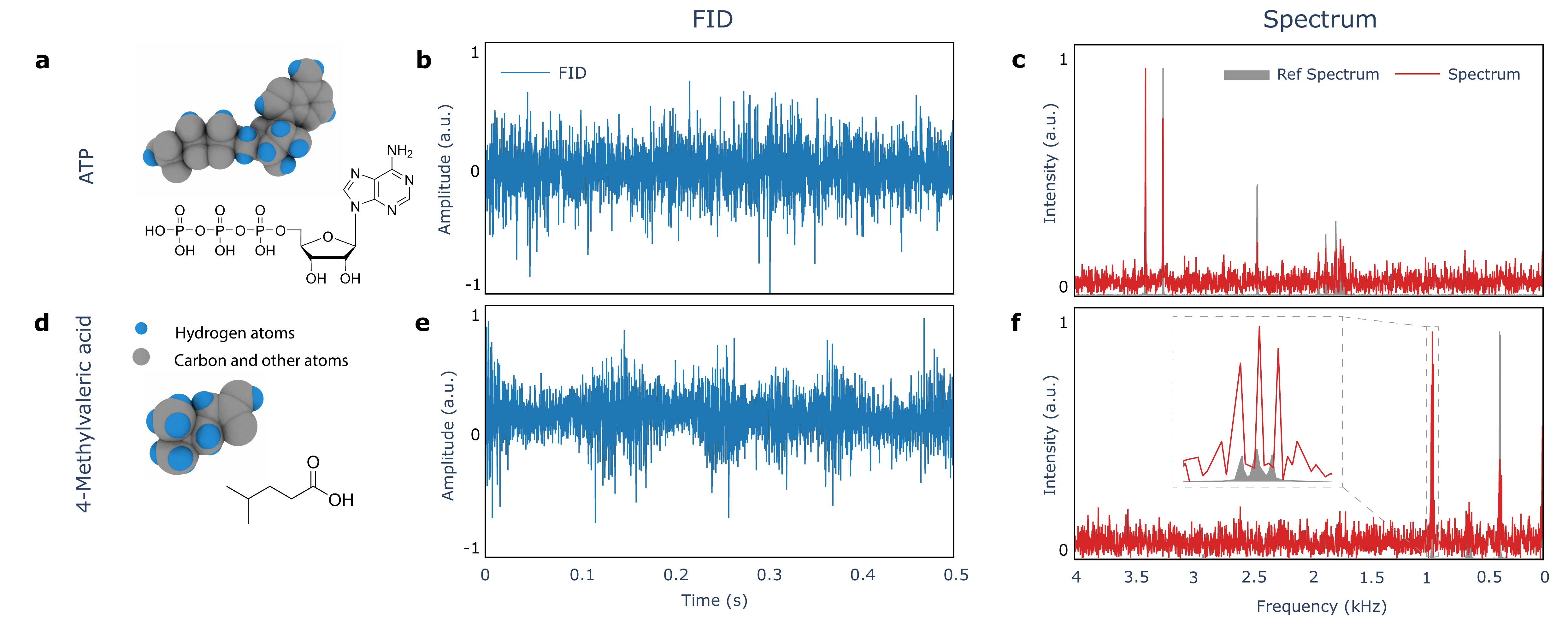}
    \caption{Shows the $\qty{400}{\MHz}$ FID signals and spectra of two spin clusters obtained from IBM Eagle r3 "brisbane" quantum device (red). The spin clusters are (a, b, c) ATP ($8$ spins, low connectivity) and (d, e, f) 4-methylvaleric acid ($11$ spins, high connectivity). The spectra show correspondence to the reference spectra obtained experimentally (grey). In some cases, as shown in the inset in (f), the J-coupling structure can be reproduced in the simulation. Despite that, the noise floor prevents obtaining the signals from all peaks, and the artifacts from using a product formula are also visible. The FID signals consist of $4096$ temporally equidistant points. The expectation value of the $X$ magnetization operator is calculated by taking $4000$ shots of the quantum circuit. The shown FID is an average of over $5$ runs on the quantum device.}
    \label{spectra}
\end{figure*}

\section{Implementation and results}

In this work, we used one repetition of the Lie-Trotter product formula as the algorithm for Hamiltonian simulation. To our knowledge, this method requires the least amount of quantum operations - \emph{gates} - and is easy to implement, but it is also a very rough approximation. It is not common to use product formulas in this way, usually the time evolution is discretized with some time step $\delta t$. This time step is chosen by weighing the cost of introducing smaller time steps versus the benefit in the accuracy that the smaller time step provides. The reason for our choice of taking only \emph{one} time step is that the main goal of this work is to obtain NMR signals from some of the largest spin systems that can be simulated on quantum processors today. The model error stemming from the choice of the product formula is secondary to this work.

Another, quantitative justification for this approach is detailed in figure \ref{trotter}. Here we compare two product formulas, the Lie-Trotter and Suzuki-Trotter formulas. We analyzed a subset of the GISSMO database \cite{gissmo} consisting of $4$-spin molecules with non-trivial connectivity to get an idea of the scaling of the model error due to the product formula approximation. In our investigations, we have found no reason to believe that in this regime of low number of repetitions increasing the number of repetitions leads to a better model.

As a metric for the error, we used the cosine distance between the spectra approximated by a product formula and the spectra calculated from first principles by taking the exponential of the Hamiltonian. The cosine distance treats spectra as high-dimensional vectors and is defined as $1-\frac{\overline{x}\cdot\overline{y}}{\abs{\overline{x}\cdot\overline{y}}}$ for $\overline{x}$ and $\overline{y}$ - two sets of points that define the spectra. This metric was chosen because it allows us not to worry about scaling the spectra to the same magnitude, which in this case is not a well-defined operation since the spectra may have quite different shapes. The cosine distance is bound by 1 for low similarity and 0 for high similarity of spectra.

 We found that the error increases initially up to $4$ repetitions for the Lie-Trotter formula (fig. \ref{trotter} \textbf{a}) and up to the fourth order for the Suzuki-Trotter formula (fig. \ref{trotter} \textbf{b}) for the considered subset of the GISSMO database. Given that the number of required gates and the gate depth grows linearly with the number of repetitions (fig. \ref{trotter} \textbf{a}), and we already are operating at the limits of the NISQ hardware available today, we concluded that in the current regime, it makes sense to limit the number of Trotter repetitions to just one. Recent results on optimizing Trotter parameters could lead to better results in the future \cite{PhysRevResearch.5.043035, PhysRevResearch.5.023146, trenkwalder2023compilation}.

\paragraph{\textbf{Product formula.}}

The order of gates in the product formula is chosen as follows:

\begin{equation}\label{pf}
    \begin{split}
    e^{-it(\sum^N_{k=1} \omega_{k}I^Z_k + 2\pi\sum_{k<l}J_{kl}\overline{I}_k\overline{I}_l)} =
    e^{-2it\pi\sum_{k<l}J_{kl}I^Z_kI^Z_l} *\\
    e^{-it\sum^N_{k=1} \omega_{k}I^Z_k} *\\
    e^{-2it\pi\sum_{k<l}J_{kl}I^Y_kI^Y_l} *\\
    e^{-2it\pi\sum_{k<l}J_{kl}I^X_kI^X_l}
    \end{split}
\end{equation}

The reasoning here is largely heuristic. If all Hamiltonian terms would commute with each other, the product formula would be not an approximation but an identity. This is not the case here, instead, the Hamiltonian terms that \emph{do} commute with each other are grouped into $4$ groups, as seen in equation \ref{pf}. Additionally, $\sum^N_{k=1} \omega_{k}I^Z_k$ always commutes with $2\pi\sum_{k<l}J_{kl}I^Z_kI^Z_l$ and sometimes commutes with $2\pi\sum_{k<l}J_{kl}I^Y_kI^Y_l + 2\pi\sum_{k<l}J_{kl}I^X_kI^X_l$, see equation \ref{commutation}.

\paragraph{\textbf{Gate counts - quadratic scaling.}}

The size of the quantum circuit required to implement this product formula is displayed in figure \ref{gates}. The circuits were transpiled to IBM's $127$ qubit architecture quantum chip and optimized with a standard Qiskit \cite{Qiskit} pipeline, which includes some basic optimizations such as combining sequences of one-qubit gates into one gate and rewriting rules based on commutation relations of quantum operations.

The chemical shift and spin-spin coupling data was taken from the GISSMO \cite{gissmo} database. It consists of $1324$ entries of molecules, their ($^1$H) spin-spin coupling matrices, and experimental and simulated spectral data. The spin clusters present in this database are mostly in the range of up to $20$ spins.

The calculation in figure \ref{gates} is done based on the complete GISSMO database, the box plots show the distribution of required gate counts for molecules consisting of up to $20$ spins, and several more data points are available for larger molecules. The dispersions in values for gate counts come from the fact that the structure of the quantum circuit needed for FID simulation depends on the structure of the simulated molecule, and there are many molecules with the same number of spins in the GISSMO database, for instance, there are $93$ molecules with $10$ proton spins. The dispersion from different compilations of the same quantum circuit is negligible. The higher the connectivity of the spin cluster of the molecule, the higher the depth of the needed quantum circuit. The presented numbers of gates are calculated for the IBM "brisbane" quantum device (Eagle r3 architecture).

The two-qubit gate requirements for the NMR FID simulation for molecules from the GISSMO database were analyzed. Figure \ref{gates} \textbf{a} shows the total number and depth of two-qubit echoed cross-resonance (ECR) gates needed to simulate a single repetition of a product formula for a given number of spins. A quadratic fit was used, matching the data quite well. We do expect the quadratic behavior for the total number of two-qubit gates: it is proportional to the number of off-diagonal elements in the spin-spin coupling matrix, which is quadratic in the number of spins $N$.

The classically intractable regime starts as early as $N=20$. From the figure \ref{gates} we can see that in this regime the expected depth of two-qubit gates is around $100$, with more connected spin clusters requiring larger gate depth, potentially up to $300$ gates. The successful execution of quantum circuits with this number of gates is feasible to expect in the next few years. The larger instance of the two simulated in this work has the two-qubit gate depth around $150$ with a small deviation in the order $10$. In our investigations, we found that no significant signal can be achieved reliably for larger instances. Given the estimated quadratic scaling of the quantum resources needed, we believe that the simulations of Heisenberg Hamiltonian that are inaccessible via classical computation have a good potential to be solvable on quantum devices when quantum devices can execute circuits of depth in the order of $1000$ two-qubit gates.  

It is worth mentioning that while the regime relevant for quantum utility starts at around $20$ (nuclear) spins, the actual molecules are often much bigger than $20$ atoms. Examples of molecules with the average number of atoms for several indicated spin counts are given in figure \ref{gates} \textbf{b} as a reference. This shows that molecules chemically relevant for useful NMR applications could be simulated already in the NISQ era.

\paragraph{\textbf{Classical exponential barrier.}}

As a guide to the eye, an estimation of classical computation times is given in figure \ref{gates} \textbf{a} in grey. This estimation was made by a straightforward no-tricks exponentiation of the Hamiltonian matrix on an Intel i7-1165G7 2.80GHz computer. It is important to keep in mind that for the presented cases the simulation can be done much faster with many spin dynamics simulation packages, for example with SPINACH \cite{SPINACH_HOGBEN, kuprovspin}. But the exponential scaling still holds for large and highly connected spin clusters - for these cases numerical methods fail.

The gate counts in figure \ref{gates} were estimated by optimizing the circuits using the same pipeline also used for optimizing and transpiling the circuits to the quantum hardware for the presented QPU simulations.

\paragraph{\textbf{QPU experiment.}}

In this work, we have simulated the NMR spectra of two spin clusters on a QPU as shown in figure \ref{spectra}. In the circuit model, the qubits are usually initialized to point along the $Z$ axis. For the sake of this work, we assume this also to be the case here, which means that as a first step, we need to initialize the qubit register to correspond to the state of the spins after the $\frac{\pi}{2}$ pulse. This step can be skipped by relabeling the coordinate system in such a way as to exchange the $X$ and $Y$ axes. Then we can assume that the spins point along the $X$ axis immediately after initialization. We kept this step in to keep the correspondence to the actual NMR experiment since it only changes the gate depth by one.

After that, the gates that simulate the interactions of spins with the field and between each other are added according to the chosen product formula. Finally, the transverse magnetization operator $\sum^N_{k=1} I^X_k$ is measured along any direction in the transverse plane, we chose the $X$ direction.

The mapping can be seen schematically in figure \ref{circuit} \textbf{d} for the example of 3-Amino-5-ethyl-1H-pyrazole.

The quantum circuits for the experiments presented in this work were run on the IBM Eagle r3 "brisbane" quantum chip. It consists of $127$ fixed-frequency transmon qubits with heavy-hex connectivity and median $T_1=222\mu s$ and $T_2=145\mu s$.

The FID signal (fig. \ref{spectra} \textbf{b} and \textbf{e}) is obtained by taking $4096$ equidistant points of time, calculating the evolution of the quantum state, and measuring the expectation value of the total $X$ magnetization operator for each of those points. The expectation value is calculated by taking $4000$ shots. The FID signal is time-averaged over $5$ such runs on the quantum hardware. Overall, repeating the experiment $N$ times, one expects an improvement in the signal-to-noise ratio by a factor of $\sqrt{N}$.

The choice of the sampling interval between the time points is guided by the range of frequencies that need to be captured for the NMR signal in the rotating frame according to the Nyquist theorem. In the case of proton NMR with the magnetic field equal to \qty{400}{\MHz} this amounts to the frequency range $[0,4000]$ Hz. This gives the sampling rate of $8000$ Hz and the sampling interval of $0.125$ ms. In NMR, the magnetic field is commonly and conveniently measured in terms of the Larmor frequency $\omega_0=\gamma B_0$ of a given nucleus type.

The spectrum (fig. \ref{spectra} \textbf{c} and \textbf{f}) is obtained by applying the Fourier transformation to the FID signal.

We have simulated the NMR spectra of two spin clusters. The larger spin cluster is the 4-methylvaleric acid \ce{C6H12O2} (fig. \ref{spectra} \textbf{a}), an organic acid with $11$ proton spins. The corresponding NMR FID simulation needs $11$ qubits and the two-gate depth in the range of $140-160$, where the depth of gates is probabilistic since the optimization algorithm for the quantum circuit is also probabilistic.

The other spin cluster is ATP (fig. \ref{spectra} \textbf{a}), containing $8$ spins. The two-qubit gate depth needed for the simulation is in the range of $35-40$. Interestingly, ATP is found in all known forms of life and is often referred to as the "molecular unit of currency" of intracellular energy transfer. Moreover with the chemical formula \ce{C10H16N5O13P3}, ATP is composed of 47 atoms, showing that already within the NISQ era, one can simulate large and relevant molecules from the standpoint of NMR.

The demonstrated simulations required around $2$ hours of quantum computing run time according to the IBM Quantum Platform web interface.

\paragraph{\textbf{Verification with NMR experimental results.}}

The practical application considered here offers the advantage of being efficiently verifiable since one can easily conduct an NMR experiment to benchmark the results from the quantum computer. For this work, we have been able to compare the spectra calculated on the QPU with real experimental spectra, thus validating this advantage towards quantum utility. The experimental spectra for ATP and 4-methylvaleric acid (fig. \ref{spectra} \textbf{c} and \textbf{f}) were obtained on a Brucker-Avance DPX FT-NMR $\qty{400}{\MHz}$ instrument. The solvents used are \ce{D2O} for ATP and \ce{CDCl3} containing $0.03\%$ Tetramethylsilane (resulting in a peak at $\qty{7.26}{ppm}$) for 4-Methylvaleric acid.

The simulated spectra show a correspondence to the experimental ground truth spectra when it comes to the general location of the peaks, artifacts often cover the finer details such as peak splittings due to the product formula approximation. In some cases, the splitting structure corresponds to the one in the experimental, as can be seen in the inset of figure \ref{spectra} \textbf{f}. Smaller peaks cannot be faithfully represented since they fall below the noise level.

\section{Conclusion}

We have demonstrated that the nuclear spin dynamics problem can benefit from the NISQ-based approaches. Our investigation into the application of NISQ devices for simulating high-field regime NMR experiments marks a significant step towards realizing quantum advantage for practical problems. The three properties of a quantum utility experiment \cite{Aaronson} were demonstrated: its in-principle quantum advantage, the possibility to implement it with NISQ resources, and crucially its efficient verifiability by benchmarking against NMR experiments. By employing a product formula with minimal resource overhead to map NMR interactions onto a QPU, we have successfully simulated liquid-state proton NMR spectra for relevant molecules with up to $11$ spins and a total of $47$ atoms. Through comparisons with real NMR experiments, despite current limitations, our findings lay the groundwork for achieving quantum utility, where computationally relevant problems can be solved by quantum computers but not by conventional means. Additionally, we provided an experimental estimation of the quantum resources required for solving larger instances of the problem using our approach. The demonstrated polynomial scaling on real processors represents a foundational step in advancing practical quantum computation towards reality. We believe that in the next few years new QPUs, combined with more advanced circuit optimization and error mitigation methods will achieve useful classically intractable simulations of NMR FID or other experiments.

\clearpage

\section{Data availability}\label{Data}

The datasets generated and analyzed during this study are available at \href{https://zenodo.org/doi/10.5281/zenodo.11069324}{Zenodo}.

\section{Acknowledgment}\label{Acknowledgment}

We acknowledge the use of IBM Quantum services for this work. The views expressed are those of the authors and do not reflect the official policy or position of IBM or the IBM Quantum team. We acknowledge Christelle Jablonski for NMR support, and Flavio Rump, Ilya Kuprov, and Julien Baglio for useful discussions. We acknowledge support from the National Centre of Competence in Research (NCCR) SPIN, funded by the Swiss National Science Foundation (grant number 51NF40-180604), and from uptownBasel. We thank QuantumBasel for the access to the IBM Quantum services and support. This research used the computational cluster resource provided by FHNW HLS.

\bibliography{bibliography}

\end{document}